\documentstyle[11pt]{article}
\newcommand{\vs}[1]{\rule[- #1 mm]{0mm}{#1 mm}}
\newcommand{\beq}{\begin{equation}}
\newcommand{\eeq}{\end{equation}}
\newcommand{\beqn}{\begin{eqnarray}}
\newcommand{\eeqn}{\end{eqnarray}}
\newcommand{\om}{\omega}
\newcommand{\al}{\alpha}
\newcommand{\ial}{i\bar{\al}}
\newcommand{\lam}{\lambda}

\newcommand{\eps}{\epsilon}
\newcommand{\Z}{{\cal Z}_\delta}
\newcommand{\M}{{\cal M}}
\newcommand{\dM}{{\cal D}M}
\newcommand{\Tr}{\mbox{Tr}}
\newcommand{\dW}{\frac{d}{dW}}
\newcommand{\Ve}{V_{ef\!f}}
\newcommand{\Vep}{\Ve^{\prime}}
\newcommand{\Vt}{\tilde{V}}
\newcommand{\dx}{\frac{dx}{dW}(p)}
\newcommand{\dy}{\frac{dy}{dW}(p)}
\newcommand{\dial}{\frac{d\ial}{dW}(p)}
\newcommand{\parW}{\frac{\partial}{\partial W}}
\newcommand{\cI}{\oint_{\cal C}\frac{d\om}{2\pi i}}
\newcommand{\Vp}{V^{\prime}}

\newcommand{\sect}[1]{\setcounter{equation}{0}\section{#1}}

\begin{document}

\begin{titlepage}

\rightline{NBI-HE-00-10}

\vs{10}

\begin{center}

{\LARGE {\bf Macroscopic and Microscopic (Non-)Universality of Compact
Support Random Matrix Theory}}\\[2cm]

{\Large \sc{G. Akemann$^{(1)}$ and G. Vernizzi$^{(2)}$}}\\[.5cm]
{\em $(1)$ Max-Planck-Institut f\"ur Kernphysik\\
Saupfercheckweg 1, D-69117 Heidelberg, Germany}\\
{\em and\\ $(2)$ The Niels Bohr Institute\\
Blegdamsvej 17, DK-2100 Copenhagen {\O}, Denmark}\\[.5cm]

\end{center} 

\vs{10}

\centerline{ {\bf Abstract}}

A random matrix model with a $\sigma$-model like constraint, the 
restricted trace ensemble (RTE), is solved in the large-$n$ limit. In the
macroscopic limit the smooth connected two-point resolvent $G(z,w)$ is found to
be non-universal, extending previous results from monomial to
arbitrary polynomial potentials. Using loop equation techniques we
give a closed though non-universal expression for $G(z,w)$, which 
extends recursively to all higher $k$-point resolvents. These findings 
are in contrast to the usual unconstrained one-matrix model. However, in the 
microscopic large-$n$ limit, which probes only 
correlations at distance of the mean level spacing,
we are able to show that the constraint does not modify the universal 
sine-law.
In the case of monomial potentials $V(M)=M^{2p}$, we provide a relation
valid for finite-$n$ between the $k$-point correlation function of 
the RTE and the unconstrained model.
In the microscopic large-$n$ limit they coincide which proves the microscopic 
universality of RTEs.

\end{titlepage}

\renewcommand{\thefootnote}{\arabic{footnote}}
\setcounter{footnote}{0}

\sect{Introduction}\label{intro}

Random matrix models enjoy a wide range of applications in physics due to
their property of being universal (for reviews see \cite{guhr,Been}). 
This property manifests itself in the
independence of correlation functions from the choice of the 
distribution function ${\cal P}(M)\sim\exp[-n\Tr V(M)]$,
where $M$ is an $n\times n$ matrix and $V$ is a polynomial. 
Different classes of universality are found depending on the way the 
large-$n$ limit is taken \cite{AJM,AA,BZ} and which part of the spectrum 
is investigated \cite{BZ,ADMN,BB,KF}.
However, not in all applications a distribution function of the above
form is realistic. Here, all the eigenvalues (=energy levels) of the
matrix $M$ are coupled through the Jacobian after diagonalization. 
There are situations as for example in applications
in Nuclear Physics, where in contrast to that 
the Hamiltonian of the model couples only few energy levels and is
still very well described by the above random matrix model  
\cite{guhr}. It is therefore very
interesting to study deformations and generalizations of the above
distribution ${\cal P}(M)$ and to investigate first, if the
correlations remain unchanged and second, if the property of universality 
is maintained. In this work we study a deformation which preserves the 
symmetry of the matrix model, which will be the unitary
transformations of the Hermitian matrix $M$ in our case. The symmetry
of the model is directly related to the properties of the Hamiltonian 
under rotations and time-reversal \cite{guhr}. There exists an
interesting relation \cite{ACMVII}
between the restricted trace ensembles (RTEs) 
which we will consider and the so-called Wigner ensembles \cite{Pastur,danna}, 
where different matrix elements are weighted with different
distribution functions, without being invariant under unitary 
transformations.

It has been known only quite recently
that there exist symmetry preserving 
deformations of the distribution function that
destroy the property of
universality \cite{ACMVII,Iso}. Two examples are the trace squared
ensembles which were originally introduced in the context
of Quantum Gravity \cite{das,CM}
and the generalized RTEs
which were introduced, in their simplest pure-quadratic form, by Bronk
and  Rosenzweig \cite{met,Bronk} in the context of Nuclear Physics. 
The non-universality does not necessarily spoil the
applicability to physical systems. In fact the deformation may be
introduced for physical reasons: indeed, trace-squared terms have been 
recognized as 
corresponding to higher order intrinsic curvature terms in the string action. 
Such terms have been added in order to cure an intrinsic instability of the 
theory related to a crumpled surface phase of the string world sheet 
\cite{das,CM,korchemsky}. Moreover, in the framework of random surfaces, 
these additional terms are interpreted as touching interaction terms
which  make the random surfaces touch each other.
The generalized RTEs permit the same graphical interpretation as they
have been shown to be a limiting case of a special trace squared
ensemble \cite{ACMV}. The RTEs are defined such that the exponential
weight  function gets
replaced by a constraint ${\cal P}_\delta(M)\sim\delta(A^2-1/n \; \Tr V(M))$ 
(or the $\theta$ Heaviside step function instead). 
By using the following representation of the $\delta$-function, 
$\delta(x)=\lim_{l\to\infty}\sqrt{\pi/l}\exp[-lx^2]$, the distribution
${\cal P}_\delta(M)$ can be easily seen to contain trace squared
terms. The relation between RTEs and the trace squared ensembles was
discussed in great detail for the spectral density in \cite{ACMV}
using saddle point techniques, and the canonical ensemble, the 
trace squared ensemble and the RTE were shown to agree to leading order. 
In \cite{Iso} also
the two-point function was derived for the trace squared ensembles 
and shown to be non-universal using 
saddle point equations. Here, we will present a scheme to calculate
general $k$-point functions of the RTEs in different large-$n$ regimes.

 Namely, one has to distinguish two different types of large-$n$ limits -
the macroscopic and microscopic limit - which may not all be affected
by the deformation of the distribution function ${\cal P}(M)$. 
In the example of generalized RTEs which we will study here we find
that this is precisely the case. While the macroscopic universality is 
destroyed by the global constraint, 
the microscopic correlations at short distances remain unchanged. 
In a sense, the canonical ensemble ${\cal P}(M)$
is replaced by its micro-canonical counterpart ${\cal P}_\delta(M)$. 
Therefore we have an
explicit model of statistical mechanics at hand, where the correlation 
functions of both ensembles can be calculated analytically
and then be compared for discrepancies.
Another remarkable property of RTEs is that
they  possess a finite support already at finite $n$. Similar to the
canonical Gaussian matrix model the RTEs allow for an explicit
calculation at finite $n$. This has been shown already in
\cite{ACMV} for the one-and two-point function and will be given here
for general $k$-point functions. Comparing the finite-$n$ and
$n\to\infty$ results the differences can be thought of as
finite-size corrections, when interpreting $n\to\infty$ as the
continuum limit. In the RTE these corrections appear in a different
way than in the canonical model, due to the finite support at
finite-$n$. Let us explain now in more detail how the large-$n$ limit can
be taken.

In the macroscopic large-$n$ limit no restrictions are made on the
distance between different eigenvalues. This leads to smooth,
universal two- and higher $k$-point correlation functions for the
canonical, unconstrained models \cite{AJM,AA}. Applications can be found in
two-dimensional Quantum Gravity \cite{Ambbook} as well as the theory
of  transport properties 
of mesoscopic wires \cite{Been}. For the generalized RTEs we have
shown in a  previous
publication together with our collaborators \cite{ACMVII}, that for a
certain class of potentials $V(M)$ the 2-point correlator is no longer 
universal. In this work we will extend these results to arbitrary
polynomial potentials (see also \cite{Gra}) 
and to all higher $k$-point resolvents. Since in 
ref. \cite{ACMVII} it was also shown that to leading order all
$k$-point resolvents of the $\delta$- and $\theta$-measure are equivalent
we will restrict ourselves here to the former one.

In contrast to that, in the microscopic large-$n$ limit correlations of 
eigenvalues $\lam,\ \mu$ at a distance of the mean level spacing
$| \lam-\mu | \sim 1/n$ are calculated. It is this kind of limit that
finds a  wide
range of applications in nuclear physics, condensed matter physics
(see e.g.  \cite{guhr})
and has initiated 
exact analytical solutions in the study of Dirac spectra in QCD \cite{SV}.
We will show that in generalized RTE with purely monomial potential 
$V(M)=M^{2p}$ (which includes of course the quadratic case for $p=1$)
the  constraint does not change
the local properties at short distances, in the sense that the
connected two-point correlator behaves according to the well-known ``sine-law''
of the Gaussian canonical ensemble \cite{BZ}. Indeed, this result
holds also  for higher $k$-point correlation functions. This does
not come as a surprise since a global constraint should not change
the local properties. We believe that the same is true for more general 
potentials and that universality holds here as well.

Consequently the present paper is split into two different parts.
Section 2 is devoted to the macroscopic large-$n$ limit where we use
loop equation techniques, closely following \cite{AMB93,A96}.
Although the loop equations are originally designed to calculate
higher orders in the $1/n$-expansion we will restrict ourselves to the 
planar limit. Since we find non-universality to leading order in all
connected correlation functions we do not calculate the likewise non-universal 
higher orders in $1/n$. The difficulty to deal with the
constraint will be treated in a similar way as the situation where the 
spectral density has a support consisting of several intervals \cite{A96}. In
these multi-band phases additional constraints have to be imposed to
make the solution unique \cite{DJ,Ju}. We will restrict
ourselves to hermitian matrices $M$ only, for 
non-hermitian matrices see \cite{Gra}.
Since only the planar solution is needed for our 
results they can be easily extended to orthogonal and
symplectic matrices using \cite{Itoi}. 
Extensions to
the complex matrix model are straightforward as well since the same
loop equation techniques exist in the literature \cite{AKM92,A97}. 
Very recently finite-$n$ results have been obtained for the eigenvalue density 
of Gaussian ensembles with real symmetric and complex matrices 
\cite{Delannay}.
In the second part
section 3 the microscopic large-$n$ limit is investigated 
for the RTE with a monomial potential $V(M)=M^{2p}$. Here, we generalize 
existing results for finite-$n$ of previous publication
\cite{ACMV}, by improving a technique already used in ref. \cite{ACMVII}. 
Rescaling variables in the microscopic limit 
and using the inverse Laplace transform we are able to match the 
connected $k$-point 
correlator to the well known ``sine-law'' behavior of the canonical
ensemble, thereby proving the microscopic RTE universality.
Let us stress that for the RTEs no orthogonal polynomial techniques 
are applicable.

\sect{The macroscopic limit: non-universality} 
\label{macro}

In order to calculate all correlation functions 
for the constrained matrix model with an arbitrary potential 
$V(M)=\sum_{j=1}^\infty \frac{g_j}{j}M^j$
we introduce an auxiliary potential $W(M)$
inside the partition function\footnote{We added a trivial factor of
$n^{2}$ inside the delta function.}
\beqn
\Z &\equiv& \int\dM \exp[-n\Tr W(M)] 
     \ \delta  \left(n^2A^{2}-n\Tr V(M)\right) \label{Zdelta}\ ,\\
&&W(M) \ \equiv\  \sum_{j=1}^\infty \frac{t_j}{j}M^j \ ,
\label{Waux}
\eeqn
where the two sets of variables $\{t_j\}$ and $(\{g_j\}, \ A)$ are
taken to be independent.
All $k$-point resolvent operators can then be obtained by taking functional
derivatives of $\Z$ with respect to $W(p)$ as given below, where we then
eventually set the auxiliary potential $W$ to zero at the end. A similar trick 
has been used in ref. \cite{Iso} in order to investigate a multi-trace 
random matrix ensemble \cite{das,CM}, showing their non-universality as well.
Furthermore we use the complex representation of the $\delta$-function
to obtain
\beq
\Z \ =\ \int \frac{d\al}{2 \pi} \int\dM 
\exp\left[-n\Tr\left(W(M)+i\al(V(M)-A^2)
\right)\right] \ .
\label{Zd}
\eeq
If we had used instead the following representation of the $\delta$-function,
$\delta(x)=\lim_{l\to\infty}\sqrt{\pi/l}\exp[-lx^2]$,
we would have obtained the trace squared ensemble:
${\cal P}_l(M)\sim\exp[2lnA^2\Tr V(M)-l\Tr V(M)^2]$, with the
strength of the touching interaction being proportional to $l$. However, it is
not straightforward to derive and solve loop equations for such an 
ensemble. In particular, we cannot directly employ the non-universality 
results of \cite{Iso}, where it was crucial that the single- and 
multi-trace potentials were {\it different}.
It is in the form eq. (\ref{Zd}) that 
we can actually derive and solve the loop equations for the constrained
model. Throughout the paper the same notation as in \cite{AMB93} is used,
which is redisplayed here for completeness.
The resolvent or 1-loop correlator is defined as 
\beq
G(p) \ \equiv \ 
       \frac{1}{n} \left\langle \Tr \frac{1}{p-M} \right\rangle_\delta 
  \ = \ \frac{1}{n} \sum_{k=0}^\infty 
             \frac{\langle \Tr M^k \rangle_\delta}{p^{k+1}} 
\label{G}
\eeq
and higher $k$-point resolvents are given by
\beq
G(p_1,\ldots,p_k) \ \equiv \ n^{k-2} \left\langle \mbox{Tr} 
       \frac{1}{p_1-M}
    \cdots \mbox{Tr} \frac{1}{p_k-M}\right\rangle_{\delta,\ conn} 
\label{Gk}
\eeq
where $conn$ stands for the connected part of the expectation value
with respect to eq. (\ref{Zd}). They are defined such that the leading part 
is of the order O(1).
If we define the free energy ${\cal F_\delta}$ as follows 
\beq
\Z \ \equiv \ \exp[n^2 {\cal F_\delta}]
\eeq
all resolvents can be obtained from it by successive applications of the
loop insertion operator
\beqn
\dW(p) &\equiv& - \sum_{j=1}^\infty \frac{j}{p^{j+1}} \frac{d}{dt_j} \ ,
\label{dW}\\
G(p_1,\ldots,p_k) &=& \dW(p_k)\dW(p_{k-1})\cdots\dW(p_1) {\cal F_\delta}
                       \ +\ \delta_{k,1}\frac{1}{p_1} \ .
\label{dGk}
\eeqn
In particular all higher resolvents can be derived from the 1-point 
resolvent alone.

\subsection{The loop equation in the planar limit}

The loop equation is derived in the usual way from the partition
function eq. (\ref{Zd}) by shifting variables $M\to\ M+\eps/(p-M)$ 
 and requiring it to be invariant under this shift, i.e. 
$\frac{d\Z}{d\eps}|_{\eps=0}=0$, 
\beq
\cI \frac{\Vep(\om)}{p-\om} G(\om) \ = \ 
   G(p)^2 + \frac{1}{n^2}\dW(p)G(p) \ ,
\label{loop}
\eeq
where we have defined the effective potential 
\beq
\Ve(M) = W(M) +\ial V(M) \ .
\label{Veff}
\eeq
In eq.(\ref{loop}) the integration contour ${\cal C}$ encircles the 
support $[y,x]$ of the spectral density $\rho(\lam)$
counterclockwise in the complex plane, not including the argument
$p\notin[y,x]$. 
The parameter $\ial$ inside the effective potential $\Ve$ is determined 
by the constraint $\langle \Tr V(M)\rangle=nA^2$  as a function of all
coupling constants, as we will see in more detail below.
This constraint can also obtained by
requiring the invariance of ${{\cal Z}_\delta}$ under the shift 
$\alpha\to \alpha+\eps$.

Let us stress again that the resolvents are given by differentiating
with respect to $W(p)$ and not the effective potential $\Ve(p)$.
Due to the $\al$-integral in the partition function $\Z$ eq. (\ref{Zd})
we have $\langle \Tr M^k\rangle_\delta\neq \langle i\al\Tr M^k\rangle_\delta$ 
which is needed to determine $G(p)$ eq. (\ref{G}). 
Furthermore let us note that
eq. (\ref{loop}) looks almost identical to the loop equation of the 
unconstrained hermitian matrix model \cite{AMB93} defined in the next section
eq. (\ref{mic1}).
However, the role of the above mentioned auxiliary potential $W$ as well
as the constraint will modify the results of \cite{AMB93}. The constraint
leads to similar complications as in the 
situation where the support of the spectral density consists of 
several intervals \cite{A96}.

In order to solve the loop equation we introduce a $1/n^2$ expansion
for all $k$-point resolvent operators (\ref{Gk})
\beq
G(p_1,\ldots,p_k) \ =\ \sum_{g=0}^\infty \frac{1}{n^{2g}} G_g(p_1,\ldots,p_k)
\ , 
\label{Gk1/n}
\eeq
where the leading part with genus $g=0$ (planar) is of the order O(1).
In ref. \cite{ACMVII} it was shown that for monomial potentials the
expectation values of the RTEs possess such an  expansion in $1/n^2$ as well.
Here we assume that the same holds true for all polynomials potentials.
Inserting this expansion into the loop equation (\ref{loop})
and taking the large-$n$ limit we obtain to leading order
\beq
\cI \frac{\Ve^{\prime}(\om)}{p-\om} G_0(\om) \ = \ G_0(p)^2 \ . 
\label{plan} 
\eeq
If we make the Ansatz that $G_0(p)$ has just one cut in the complex
plane or equivalently the support of the eigenvalues consists of the
single interval $[y,x]$ we obtain
\beq
G_0(p) \ = \ \frac{1}{2}\left( \Vep(p)-\M(p)
            \sqrt{(p-x)(p-y)}\right) \ , 
\label{ansatz} 
\eeq
where the analytic function $\M(p)$ is given by
\beq
\M(p) \ = \ \oint_{\cal C_{\infty}} \frac{d\om}{2\pi i}
\frac{\Vep(\om)}{(\om-p)\sqrt{(\om-x)(\om-y)}}
\label{M} \ .
\eeq
For details of the derivation see for example ref. \cite{A96}.
The final result can be written as follows, 
after deforming back the integration contour,
\beq
G_0(p) \ =\  \frac{1}{2}\cI \frac{\Vep(\om)}{p-\om}
           \sqrt{\frac{(p-x)(p-y)}{(\om-x)(\om-y)}} \ .
\label{G0}
\eeq
To make the solution complete we still have to determine the endpoints 
$x$ and $y$ as well as the parameter $\ial$ in terms of the coupling
constants of $W$ and $V$ and the parameter $A$. The first two
equations can be obtained from the asymptotic behavior of
$G_0(p)$. According to the definition (\ref{G}) we have
\beq
\lim_{p\to\infty}G(p) \ = \ \frac{1}{p} \ . 
\label{Gass}
\eeq
Since the leading term does not depend on $n$ it 
comes from the planar part $G_0(p)$ and we obtain the conditions
\beqn
0&=& \frac{1}{2} 
 \cI \frac{\Vep(\om)}{\sqrt{(\om-x)(\om-y)}} \ ,\nonumber\\
1&=& \frac{1}{2} 
 \cI \frac{\om \Ve^{\prime}(\om)}{\sqrt{(\om-x)(\om-y)}} \ .
\label{bc1}
\eeqn
The third equation needed we obtain from the constraint on $\Tr V(M)$
which we rewrite in terms of the spectral density
\beqn
\rho(\lam)\ &=& \ \frac{1}{2\pi i}\lim_{\eps\to 0}
         \Big( G_0(\lam-i\eps)-G_0(\lam+i\eps) \Big) \nonumber\\ 
          \ &=& \ \frac{1}{2\pi}\M(\lam)\sqrt{(x-\lam)(\lam-y)} \ . 
\label{rho}
\eeqn
The constraint then reads
\beq
A^2 \ =\ \int_y^x \!d\lam\ \rho(\lam) V(\lam) 
    \ =\ \int_y^x \!\frac{d\lam}{2\pi}\M(\lam)
    \sqrt{(x-\lam)(\lam-y)}\ V(\lam)
\label{bc2} \ ,  
\eeq
which determines the parameter $\ial$ contained in $\M(\lam)$. This
equation together with the boundary conditions eq. (\ref{bc1})
determines the planar resolvent $G_0(p)$ eq. (\ref{G}) completely as a
function of the coupling constants of $W$ and $V$ and of the parameter 
$A$.

\subsection{Higher planar $k$-point resolvents}

Starting from the the planar resolvent $G_0(p)$ we can obtain all
higher planar $k$-point resolvents by successively applying the loop
insertion operator $\dW$ to it, as it is given in
eq. (\ref{dGk}). Here we use the fact that all resolvents 
have the same expansion in $1/n^2$ eq. (\ref{Gk1/n}). 
For this purpose we introduce a set of new parameters $M_k$ and $J_k$, 
$k\in$ N$_+$.  These moments usually
play the role of universal parameters of the higher $k$-point
resolvents encoding all the information of the potential in addition
to the endpoints $x$ and $y$ \cite{AMB93}. We will then rewrite the
loop insertion operator in terms of these new variables. This is done
in order to make the successive application of $\dW$ to an algebraic
procedure. Finally we calculate explicitly the non-universal planar 2-point
resolvent $G_0(p,q)$ and comment on the general situation.

Let us begin by defining
\beqn
M_k &\equiv& \cI \Vep(\om)\frac{\phi(\om) }{(\om-x)^k}
\ ,\ \ \ \ \phi(\om)\ \equiv\ \frac{1}{\sqrt{(\om-x)(\om-y)}} \ ,\nonumber\\
J_k &\equiv& \cI \Vep(\om)\frac{\phi(\om) }{(\om-y)^k}
\ ,\ \ \ \ k=1,2,\ldots \ .
\label{moments}
\eeqn
Expanding the poles at $x$ and $y$
the moments can be explicitly written as functions of the
coupling constants. Because of
$M_k=\frac{1}{(k-1)!}\frac{d^{k-1}}{d\lam^{k-1}}
         M(\lam)\Big |_{\lam=x}$ and similarly for $y$
the moments also characterize the multi-critical points of the model.
We now rewrite the loop insertion operator eq. (\ref{dW}) in the
following way 
\beqn
\dW (p) \ &=&\ \parW (p)\ +\ \dx \frac{\partial}{\partial x} 
                        \ +\ \dy \frac{\partial}{\partial y}
                        \ +\ \dial \frac{\partial}{\partial \ial} \nonumber\\  
 &&+\sum_{k=1}^{\infty}\left(\frac{dM_k}{dW}(p)\frac{\partial}{\partial M_k}  
    \ +\ \frac{dJ_k}{dW}(p)\frac{\partial}{\partial J_k}  \right)\ ,
\label{dWsum}\\
\parW (p) \ &\equiv & \ - \sum_{j=1}^\infty \frac{j}{p^{j+1}} 
                      \frac{\partial}{\partial t_j} \ . 
\eeqn
While the $\frac{dM_k}{dV}(p)$ and $\frac{dJ_k}{dV}(p)$ can be
obtained in a straightforward way from the definition
eq. (\ref{moments}) (see e.g. \cite{AMB93}) the remaining unknown
quantities are derived by applying $\dW (p)$ to the boundary conditions
eqs. (\ref{bc1}) and (\ref{bc2}) and solving a linear set of
equations. This is done in the Appendix \ref{detxya} with the result
reading
\newpage
\beqn
M_1\dx &=& \frac{\phi(p)}{p-x} \ +\ \frac{4}{B(x-y)}(G_0(p)-\phi(p)) \ ,
\nonumber\\
J_1\dy &=& \frac{\phi(p)}{p-y} \ -\ \frac{4}{B(x-y)}(G_0(p)-\phi(p))\ ,
\nonumber\\
\frac{1}{\ial}\dial &=& -\frac{1}{B}(G_0(p)-\phi(p)) \ ,
\label{dxya}
\eeqn
where
\beqn
\label{B}
B &\equiv& \ial \int_y^x\! d\lam V(\lam)
 \frac{1}{2\pi i}\lim_{\eps\to 0} \times \\
 &&\times\ 
         \Big[ G_0(\lam-i\eps)-\phi(\lam-i\eps)
           \ -\ (G_0(\lam+i\eps)-\phi(\lam+i\eps)) \Big]. \nonumber
\eeqn
All quantities are expressed by elementary functions and the
planar resolvent $G_0(p)$ eq. (\ref{G0}) and we have set already the
auxiliary potential $W\equiv 0$. The result for general $W$ can be derived
from eqs. (\ref{dbc1}) and (\ref{dbc2}).

We are now ready to apply the loop insertion operator $\dW$ in the form
(\ref{dWsum}) to $G_0(p)$ eq. (\ref{G0}), which does not explicitly
depend on the moments:
\beqn
G_0(p,q) &=& \dW(q)G_0(p) \nonumber\\
         &=&\frac{1}{2(p-q)^2}\left( \frac{\phi(q)}{\phi(p)} -1\right)
 + \frac{1}{4(q-p)}\frac{\phi(q)}{\phi(p)}
    \left( \frac{1}{q-x}+\frac{1}{q-y}\right) \nonumber\\
&&+ \frac{d\ial}{dW}(q)\frac{1}{2} \cI \frac{\Vp(\om)}{p-\om}
           \sqrt{\frac{(p-x)(p-y)}{(\om-x)(\om-y)}} \nonumber\\
&&+ \frac{1}{4\phi(p)}\left( \frac{1}{p-x}M_1\frac{dx}{dW}(q)
\ +\ \frac{1}{p-y}J_1\frac{dy}{dW}(q)\right) \ ,
\eeqn
after performing some contour integrals.
If we set $W\equiv 0$ we can use the results eq. (\ref{dxya}) and
finally obtain 
\beq
G_0(p,q) \ =\ G_0^{can}(p,q) \ -\ 
\frac{1}{B}(G_0(p)-\phi(p))(G_0(q)-\phi(q)) \ ,
\label{G0pq}
\eeq
the planar connected 2-point resolvent of the constrained matrix
model. The first part is the well known universal 2-point resolvent
of the corresponding unconstrained or canonical matrix model ({\it can})
\cite{AJM} 
\beq
G_0^{can}(p,q)\ =\ 
\frac{1}{4(p-q)^2}\left( \frac{(p-x)(q-y)+(p-y)(q-x)}
{\sqrt{(p-x)(p-y)(q-x)(q-y)}}-2 \right)
\eeq
whereas the second part contains the non-universal terms $G_0(p)$ and 
$G_0(q)$. Still, the result is given in 
closed form for an arbitrary polynomial potential $V(M)$. 
Eq. (\ref{G0pq}) can be compared with the earlier result \cite{ACMVII} 
\beq
G_0(p,q) \ =\ G_0^{can}(p,q) \ -\ \frac{1}{2p}\partial_p\left(pG_0(p)\right)
\partial_q\left(qG_0(q)\right)
\label{G0pqold}
\eeq
for the special case of monomial potentials $V(M)= M^{2p}$, $p\in$ N$_+$. 
For $p=1,2$ we have checked explicitly that the  two results 
eq. (\ref{G0pq}) and eq. (\ref{G0pqold}) agree. The corresponding resolvents
can be found in \cite{ACMVII}\footnote{In eq. (2.18) 
of ref. \cite{ACMVII} the factor of $1/2p$ is missing.}  
and the parameter $\ial$ was already determined in \cite{ACMV}. 
From the procedure described above it is clear that also the higher
$k$-point resolvents will remain non-universal since the derivative 
$\dW(p_i)G_0(p_j)$ always contains terms proportional to $G_0(p_i)$ and
$G_0(p_j)$.

Let us also briefly comment on higher genus contributions. Expanding
the loop equation in $1/n^2$ together with eq. (\ref{Gk1/n}) one obtains
for genus one
\beq
\left(\hat{\cal K}-2G_0(p)\right)G_1(p) \ = \ G_0(p,p) \ , 
\label{loop1} 
\eeq
where
\beq
\hat{\cal K} f(p) \ \equiv \ \cI \frac{\Vep(\om)}{p-\om}f(\om) \ .
\eeq
The right hand side of eq. (\ref{loop1}) is easily obtained from
eq. (\ref{G0pq}) by setting $p\!=\!q$. However, it is no longer a 
rational function
in contrast to $G_0^{can}(p,p)$. Its non-universality will then
translate to $G_1(p)$ after inverting the integral operator 
$(\hat{\cal K}-2G_0(p))$ and thus to higher genera through
\beq
\left(\hat{\cal K}-2G_0(p)\right)
G_g(p) \ = \ \sum_{g\prime=1}^{g-1}G_{g\prime}(p)
          G_{g-g\prime}(p)+\dW(p)G_{g-1}(p) \ , \ \ g\ge 1 \ .
\label{loopg}
\eeq
For this reason we do not go through the tedious procedure of finding
a basis for $(\hat{\cal K}-2G_0(p))$ now including also square 
roots and inverting it.

\newpage
\sect{The microscopic limit: universality} 
\label{micro}

In this section we investigate correlations of eigenvalues at the distance of
the mean level spacing $D\sim 1/n$, the so-called microscopic large-$n$ limit.
We will heavily exploit the knowledge about correlations in the
unconstrained canonical model for finite as well as infinite $n$. Our
main  result is an
explicit relation between the canonical and RTE correlations for finite-$n$
which then serves to determine the microscopic RTE correlations and prove
their universality.

Let us recall the known results about the canonical 
ensemble which also fixes our notation. The partition function reads
\beq
\label{mic1}
{\cal Z} \equiv \int \dM \; \exp [-n \Tr \Vt(M)] \  \ ,
\eeq
where $\Vt(M)$ is a polynomial. For the generalized RTE or micro-canonical 
ensemble
\beq
\label{Zdel}
{\cal Z}_\delta \equiv \int \dM \;  \delta  \left(A^{2}-\frac{1}{n} 
\Tr V(M)\right)\ ,
\eeq
where we do not need to introduce an auxiliary potential 
in contrast to the previous section. 
In order to obtain the same microscopic correlations for the two
models we will eventually have to relate the coupling constants of the 
respective polynomial potentials $V$ and $\Vt$ (as in the macroscopic limit, 
see \cite{ACMV}). The $k$-point density correlation function is defined as
\beq
\label{mic5}
\rho(\lam_1,\ldots,\lam_k) \equiv 
 \left\langle \frac{1}{n}\mbox{Tr}\; 
       \delta(\lam_1-M)
    \cdots \frac{1}{n}\mbox{Tr} \;\delta({\lam_k-M})\right\rangle \ ,
\eeq
and similarly for the delta-measure. Its connected part $conn$ is related
in the following way to the $k$-point resolvent defined in the previous 
section eq. (\ref{Gk})
\beqn
\label{mic4}
\rho^c(\lam_1,\ldots,\lam_k) &\equiv& 
 \left\langle \frac{1}{n}\mbox{Tr}\;        \delta(\lam_1-M)
   \cdots \frac{1}{n}\mbox{Tr} \;\delta({\lam_k-M})\right\rangle_{conn},\\
&=&\frac{1}{n^{2k-2}} 
\left( \frac{-1}{2\pi i} \right)^{k} \times \nonumber\\
&&\times\lim_{\eps\to 0}
\sum_{\sigma_i=\pm} \left( \prod_i \sigma_i \right) 
G(\lam_1+\sigma_1i\eps,\ldots,
\lam_k+\sigma_ki\eps) \ , \nonumber
\eeqn
where we have not yet taken the large-$n$ limit.
In this section we deal with density correlations instead of resolvents 
because they can be given more explicitly for finite-$n$. Namely 
in the canonical model all density correlators can be expressed 
in terms of the Kernel $K_n(\lam,\mu)$ of a set of orthonormal polynomials
$P_l(\lam)$ at finite-$n$ \cite{Mehta} 
\beqn
\label{mic8a}
\rho(\lam_1,\ldots,\lam_k) & = & \det_{1\leq i,j\leq k} 
\left[ K_n(\lam_i,\lam_j)\right] \ ,    \\
\rho^c(\lam_1,\ldots,\lam_k) & = & (-1)^{k+1} \sum_{P} K_{n}
(\lam_1,\lam_2)  K_{n}(\lam_2,\lam_3) \cdots  K_{n}(\lam_k,\lam_1) 
\nonumber \ ,
\eeqn
where the sum is taken over the $(k-1)!$ distinct cyclic permutation
$P$ of  the indices $(1,2,\ldots,k)$. The kernel and the polynomials are
defined as
\beqn
\label{mic9}
K_n(\lam,\mu) &=& \frac{1}{n}e^{-\frac{n}{2}(\Vt(\lam)+\Vt(\mu))}
\sum_{k=0}^{n-1} P_k (\lam) P_k(\mu) \ ,\\
\delta_{kl} & = & \int d \lam  e^{-n\Vt(\lam)} 
P_k (\lam) P_l(\lam) \nonumber \ .
\eeqn
The use of the orthogonal polynomial method is only possible for the canonical
model because the measure $\exp [-n \Tr \Vt(M)]$ inside the  partition function
eq. (\ref{mic1}) factorizes in terms of the eigenvalues $\lam_i$ of the 
hermitian matrix $M$. For the RTE no such property holds which forces us to 
seek for other methods. Here we will make use of homogeneity properties 
for monomial potentials.

Let us finally give the universal results for the canonical ensemble in the
microscopic large-$n$ limit, where we restrict ourselves to the origin
scaling limit due to the local translational invariance of the canonical
ensemble. As mentioned in the beginning we measure eigenvalues in units of the 
mean level spacing which is $D=1/(n\rho(0))$ at the origin. Here 
$\rho(0)$ is the mean eigenvalue density at zero, taken in the 
{\it macroscopic} large-$n$ limit as given in eq. (\ref{rho}).
We then define new variables $z_i=\lam_i/D$ which are kept fixed
in the large-$n$ limit. Since $D\to 0$ as $n\to\infty$ 
the variables $\lam_i$ have to go to zero as well. 
In this particular limit the microscopic correlators are defined 
as\footnote{The factor $n^k$ appears as we had already defined the $k$-point
correlator in eq. (\ref{mic5}) to be normalized to unity. The appropriate 
unfolding procedure is usually defined for un-normalized correlators (see e.g.
\cite{guhr}), which provides us with the correct pre-factor $1/\rho(0)^k$.}
\beq
\label{mic6}
\rho_S(z_1,\ldots,z_k)\ \equiv\ \lim_{n \to \infty} (Dn)^k
\rho(z_1 D,\ldots,z_k D) \ .
\eeq
This limit, which is well behaved and finite, can be investigated by
using the Darboux-Christoffel formula for the kernel eq. (\ref{mic9})
and the asymptotic large-$n$ 
behavior of the polynomials $P_k(\lam)$ to obtain \cite{BZ}
\beq
\label{sin}
\lim_{n\to\infty}Dn 
K_n(z_1D, z_2D) \ =\ \frac{\sin(\pi(z_1-z_2))}{\pi(z_1-z_2)}
\ .
\eeq
This is the universal sine-law which is valid for all polynomial
potentials $\Vt(\lam)$. Together with eqs. (\ref{mic8a}) and
(\ref{mic6}) it completely determines all $k$-point density correlators,
connected and not-connected, where at coinciding arguments we have
$\lim_{n\to\infty}Dn K_n(zD,zD)=1$,
\beq
\label{rhoS}
\rho_S(z_1,\ldots,z_k)\ =\ \det_{1\leq i,j\leq k} 
\left[ \frac{\sin(\pi(z_i-z_j))}{\pi(z_i-z_j)}\right] \ ,
\eeq
and similarly for $\rho_S^c(z_1,\ldots,z_k)$.

We conclude with the following
remark. If we had taken the macroscopic large-$n$ limit instead, the 
connected correlators $\rho^c(z_1,\ldots,z_k)$
had been of the order O$(1/n^{2k-2})$ as one can see from eq. (\ref{mic4})
together with the fact that the (connected) resolvents are of order O(1).
However, when taking the microscopic limit keeping the $z_i$ fixed, 
the asymptotic kernel eq. (\ref{sin}) is of 
order O(1) and hence are the connected {\it and} not-connected microscopic
$k$-point correlators from eq. (\ref{mic8a}). 
Consequently, the knowledge of both connected and not-connected 
correlators, is equivalent here
since they can be obtained from each other by
adding or subtracting $l$-point correlators $(l<k)$ of order O(1).
In that sense the microscopic large-$n$ limit modifies the usual
large-$n$ factorization of correlation functions.

\subsection{Microscopic $k$-point correlation functions}

As it has been mentioned already the correlation functions in the
RTE  can not be calculated using orthogonal polynomials
because of the Dirac $\delta$-function in the measure.
However, in the particular case of purely
monomial  potentials $V(M)=M^{2p}$, the evaluation of the connected
$k$-point  correlator in the microscopic limit is straightforward.
We shall exploit some homogeneity properties of this case,
which make it possible to relate ensemble averages in the monomial RTE
to ensemble averages of the same quantities in the canonical
ensemble\footnote{This technique is a slight generalization of
ref. \cite{ACMVII}, where only the macroscopic limit
was  investigated.} with potential $\Vt(M)=gM^{2p}$.
Let us define a function $F_k(M;{\bf \lam})$ of the
matrix $M$  and of a set of parameters ${\bf \lam}=(\lam_1,\ldots,\lam_k)$,  
that satisfies the following homogeneity property under a rescaling of 
the matrix
\beq
\label{hom}
F_k(tM;{\bf \lam})\ =\ t^{a}F_k(M;t^{b}{\bf \lam})\ \ \ \ 
 \mbox{for} \ \ \ \ t,a,b\in \mbox{R} \ .
\eeq
A simple example for such a function is the operator 
$\Tr \delta(\lam-M)/n$
which has $a=b=-1$. Also any such product is a homogeneous function, as 
inside the average of the $k$-point correlation function eq. (\ref{mic5}),
having $a=-k$ and $b=-1$.
In appendix \ref{appb} we derive the following formula for such homogeneous 
functions $F_k$, which relates their canonical and RTE average
\beq
\label{mic10}
\langle F_k(M;\lam) \rangle_{\delta}\ =\ 
\frac{(gn^2)^{\frac{a}{2p}}
\Gamma\left(\frac{n^2}{2p}\right)}{ A^{\frac{n^2}{p}-2}} 
\ {\cal L}^{-1} \left[  
 \frac{\langle F_k(M;\left( \frac{gn^2}{t} \right)^{\frac{b}{2p}}\lam) 
\rangle}{t^{\frac{n^2+a}{2p}}} \right] (A^2) \ .
\eeq
Here ${\cal L}^{-1}[h(t)](x)$ is the inverse Laplace transform of a
function  $h(t)$, evaluated at the point $x>0$ (for an integral 
representation see eq. 
(\ref{Ldef})). Eq. (\ref{mic10}) holds for any finite $n$. If we choose
the 1- or 2-point correlator from eq. (\ref{mic5}) as an example
we reproduce the finite-$n$ results of ref. \cite{ACMV} for 
$\rho_\delta(\lam)$ and $\rho_\delta(\lam,\mu)$ which were given
in the case of a Gaussian potential $p=1$.
If we choose for general $k$ to take
$\langle F_k(M;\lam) \rangle_\delta= \rho_{\delta}(\lam_1,\ldots,\lam_k)$, 
which obviously 
fulfills the criterion (\ref{hom}), we obtain from eq. (\ref{mic10}) 
the following expression for the $k$-point RTE correlator
\beq
\label{rhorel}
\rho_\delta(\lam_1,\ldots,\lam_k) \ = \ 
\frac{(gn^2)^{\frac{-k}{2p}}
\Gamma\left(\frac{n^2}{2p}\right)}{ A^{\frac{n^2}{p}-2}} 
\ {\cal L}^{-1} \!\left[  
 \frac{\rho(\left(\frac{gn^2}{t} \right)^{\frac{-1}{2p}}\!\lam_1,\ldots,
\left(\frac{gn^2}{t} \right)^{\frac{-1}{2p}}\!\lam_k)}
{t^{\frac{n^2-k}{2p}}} \right] \!(A^2) 
\eeq
Now we make use of the fact that at finite $n$ 
the correlations $\rho(\lam_1,\ldots,\lam_k)$ 
from eq. (\ref{mic8a}) can be written as a polynomial in all 
variables $\lam_i$ times an exponential measure factor:
\beq
\label{general}
\rho(\lam_1,\ldots,\lam_k)\ \equiv\ e^{-ng \sum_{i=1}^{k} \lam_i^{2p}} 
\sum_{l_1,\ldots,\l_k=0}^{2n-2}
 c_{ \{ l_1,\ldots,l_k \} }^{(n)} \lam_1^{l_1} \cdots \lam_k^{l_k}\ .
\eeq
Due to this fact we can actually perform the inverse Laplace transformation
\newpage
\beqn
\rho_{\delta}(\lam_1,\ldots,\lam_k)
   & = & \frac{\Gamma\left( \frac{n^2}{2p} \right) }
{(gn^2)^{\frac{k}{2p}} A^{\frac{n^2}{p}-2}} 
\sum_{l_1,\ldots,l_k=0}^{2n-2}
 \frac{c_{ \{ l_1,\ldots,l_k \} }^{(n)}}{(gn^2)^{\frac{1}{2p}\Sigma_i l_i}}
 \lam_1^{l_1} 
\cdots \lam_k^{l_k} \times \nonumber \\
\ && \times \ {\cal L}^{-1} \left[  
 t^{-\frac{n^2-k-\Sigma_i l_i}{2p}} \right] (A^2 -\frac{1}{n}\sum_{i=1}^{k} 
\lam_i^{2p} ) \nonumber \\
\ & = & \frac{\theta \left( A^2 -\frac{1}{n} \sum_{i=1}^{k}\lam_i^{2p} 
\right) }{(gn^2)^{\frac{k}{2p}} A^{\frac{n^2}{p}-2}} 
\sum_{l_1,\ldots,l_k=0}^{2n-2}
 \frac{c_{\{ l_1,\ldots,l_k \} }^{(n)} \lam_1^{l_1} \cdots 
\lam_k^{l_k}}{(gn^2)^{\frac{1}{2p}\Sigma_i l_i}} 
 \times \nonumber \\
\ && \times
\frac{\Gamma\left( \frac{n^2}{2p} \right)}{\Gamma \left(
    \frac{n^2-k-\Sigma_i l_i}{2p} \right)} 
\left( A^2 -\frac{1}{n} \sum_{i=1}^{k} \lam_i^{2p}
\right)^{\frac{n^2-k-\Sigma_i l_i}{2p}-1} ,
 \label{heavy}  
\eeqn
where we have used the shift property 
${\cal L}^{-1}[h(t) e^{\sigma t}](x)={\cal L}^{-1}[h(t)](x+\sigma)$,
the linearity of the inverse Laplace transform and eq.
(\ref{useful}) of Appendix B. Eq. (\ref{heavy}) is our first result, 
the finite-$n$ $k$-point correlation function for RTEs with monomial 
potential $V(M)=M^{2p}$ in terms of the corresponding canonical correlator 
at finite-$n$ with potential $\Vt(M)=gM^{2p}$.
In general, eqs. (\ref{general}) and (\ref{heavy}) are different from each
other at finite-$n$. This remains true in the macroscopic
large-$n$  limit, as we have seen in the previous section. 
In the remaining part we will show that in the microscopic large-$n$ limit,
however, they happen to coincide.

In a first step, we determine 
the mean level spacing $D_\delta=1/(n\rho_\delta(0))$ 
in order to define the appropriate
microscopic scaling limit. From eq. (\ref{heavy}) at $\lam=0$ one can read
off $\rho_{\delta}(0)$, since then the sum over all $l_i$ collapses.
We obtain
\beq
\rho_{\delta}(0)=\frac{c_{\{ 0 \}}^{(n)}}{(gn^2A^2)^{\frac{1}{2p}}} 
 \frac{\Gamma\left( \frac{n^2}{2p} \right) }{\Gamma \left(
     \frac{n^2-1}{2p}  
\right)} \longrightarrow \frac{c_{\{0 \} }}{ (2pgA^2)^{\frac{1}{2p}}} 
\eeq
for its large-$n$ value, with 
$c_{\{ 0 \} }=\rho(0)$ being the macroscopic large-$n$ limit of the 
canonical spectral density eq. (\ref{rho}) at the origin 
(which exists and is  finite). In order to have the same mean level spacing 
as in the canonical ensemble we would have to set $2pgA^2=1$. 
This identification of coupling constants occurs also in the macroscopic 
large-$n$ limit in order to match the corresponding macroscopic spectral 
densities (see refs. \cite{ACMVII,ACMV}). 
However, since we measure all correlations in units of $D$ and $D_\delta$
respectively we do not need to identify $D=D_\delta$ since they drop
out in the microscopic correlators anyway, as we will see below.

We will now take the microscopic limit analogue to eq. (\ref{mic6})
with rescaling $\lam_i=z_iD_\delta$ of our finite-$n$ relation (\ref{heavy}). 
The large-$n$ limit of the different factors can be obtained as follows,
starting with the  $\theta$-function term in eq. (\ref{heavy}) 
\beq
\theta \left( A^2 -\frac{2pgA^2}{n^{2p+1} }\sum_{i=1}^{k} 
\left(\frac{z_i}{c_{\{ 0\} }}\right)^{2p} \right) 
\longrightarrow \ \theta (A^2)=1 \ \ .
\eeq
The remaining $A$-dependent terms yield
\beqn
&&A^{-\frac{n^2}{p}+2} \left( A^2 -\frac{2pgA^2}{n^{2p+1}} 
\sum_{i=1}^{k} \left(\frac{z_i}{c_{ \{ 0 \} }}\right)^{2p} \right)^
{\frac{n^2-k-\Sigma_i l_i}{2p}-1}\ = \nonumber\\
&&=\ A^{-\frac{1}{p}(k+\Sigma_i l_i)}\left( 1\ -\ 
\left[ \frac{n^2-k-\Sigma_i l_i}{2p}-1 \right] \frac{2pg}{n^{2p+1}} 
\sum_{i=1}^{k} \left(\frac{z_i}{c_{ \{ 0 \} }}\right)^{2p}\ +\ \ldots\right)
\nonumber\\
&&=\ A^{-\frac{1}{p}(k+\Sigma_i l_i)} 
\left( 1+ \mbox{O}(\frac{1}{n^{2p-1}})\right)\ ,
\label{Aterm}
\eeqn
which still has to be evaluated under the sum over $l_i$'s.
Here we have used in the second step that the first factor inside the 
parenthesis is of O$(n^2)$ since 
\beq
\sum_{i=1}^k l_i \ \leq\ k(2n-2) \ ,
\label{li}
\eeq
is at most of O$(n)$. Because of $p\geq1$ the corrections are sub-leading. 
The factor containing $\Gamma$-functions we evaluate together 
with the explicit factors of $n$ in eq. (\ref{heavy})
\beqn
\label{Gamma}
&&\frac{\Gamma\left( \frac{n^2}{2p} \right) }{(gn^2)^{\frac{1}{2p}
(k+\Sigma_i l_i)}
\Gamma \left( \frac{n^2}{2p}-\frac{(k+\Sigma_i l_i)}{2p} \right)} \ = \\
&&=\ (2pg)^{-\frac{1}{2p}(k+\Sigma_i l_i)}
\left( 1-\frac{1}{4pn^2}(k+\sum_{i=1}^k l_i)(2p+k+\sum_{i=1}^k l_i)
+ \ldots\right) \ .\nonumber
\eeqn
It remains to be shown that the second term and thus higher terms in the 
expansion are sub-leading. A naive counting from eq. (\ref{li}) suggests
that this might not be the case. In order to use the microscopic results
for the canonical correlators eq. (\ref{rhoS}) we put together our results 
obtained so far
\newpage
\beqn
&&\lim_{n\to\infty}(D_\delta n)^k
\rho_{\delta}(z_1D_\delta,\ldots,z_kD_\delta) \ =\nonumber\\
&&=\lim_{n\to\infty} \frac{(2pgA^2)^{\frac{k}{2p}}}{(c_{\{0\} })^k}
\!\!\sum_{l_1,\ldots,l_k=0}^{2n-2}
\!\! c_{\{ l_1,\ldots,l_k \} }^{(n)}\! \left(\frac{z_1(2pgA^2)^{\frac{1}{2p}}}
{nc_{\{0\} }}\right)^{\!l_1} 
\!\!\!\!\cdots \left(\frac{z_k(2pgA^2)^{\frac{1}{2p}}}{nc_{\{0\} }}
\right)^{\!l_k} 
\nonumber \\
&&\times 
(2pgA^2)^{-\frac{1}{2p}(k+\Sigma_i l_i)}
\left( 1 - \frac{1}{4pn^2}(k+\sum_{i=1}^kl_i)(2p+k+\sum_{i=1}^kl_i)
+\ldots\right) \nonumber\\
&&=\lim_{n\to\infty} 
\left( 1-\frac{1}{4pn^2}(\sum_{i=1}^k\partial_{z_i}z_i)
(2p+\sum_{i=1}^k\partial_{z_i}z_i)+\ldots\right) \nonumber\\
&&\times \frac{1}{(c_{\{0\} })^k}
 \sum_{l_1,\ldots,l_k=0}^{2n-2}
 c_{\{ l_1,\ldots,l_k \} }^{(n)} \left(\frac{z_1}{nc_{\{0\} }}\right)^{l_1} 
\!\!\cdots \left(\frac{z_k}{nc_{\{0\} }}\right)^{l_k} .
\label{rhodfinal}
\eeqn
Here, the pre-factors from eqs. (\ref{Aterm}) and (\ref{Gamma}) have canceled
with the factors of $(2pgA^2)^{\frac{1}{2p}}$ from the unfolding.
Now we know from the canonical ensemble eqs. 
(\ref{general}), (\ref{mic6}) and (\ref{rhoS}) that the limit of the sum over 
the $l_i$ exists and is finite:
\beq
\rho_S(z_1,\ldots,z_k)\ = \ \lim_{n\to\infty}\frac{1}{(c_{\{0\} })^k}
 \sum_{l_1,\ldots,l_k=0}^{2n-2}
 c_{\{ l_1,\ldots,l_k \} }^{(n)} \left(\frac{z_1}{nc_{\{0\} }}\right)^{l_1} 
\!\!\cdots \left(\frac{z_k}{nc_{\{0\} }}\right)^{l_k}\ .
\eeq
Hence the term in eq. (\ref{rhodfinal}) proportional to $1/(4pn^2)$ is indeed
sub-leading and we have as a final result
\beq
\rho_{\delta,S}(z_1,\ldots,z_k)\ = \ \rho_S(z_1,\ldots,z_k) \ ,
\label{final}
\eeq
or more explicitly in terms of eq. (\ref{rhoS}). Since our derivation holds
for the RTE with an arbitrary monomial potential $V(M)=M^{2p}$,
we have not only derived all $k-$point correlation functions but also proved
their universality for the given class of potentials.
Let us finally point out that the equivalence eq. (\ref{final}) also 
holds for the corresponding connected $k-$point correlation functions.
As we have mentioned already at the end of the previous subsection, they are
of the same order in the microscopic limit and they 
can be obtained from each other
by adding or subtracting lower $l$-point correlators.

\sect{Conclusions}   

We have shown for RTEs as an example of constrained random matrix models
that in one and the same model correlation functions may exhibit universal
and non-universal behavior in different large-$n$ regimes. 
In particular, in the macroscopic large-$n$ limit all planar connected 
$k$-point resolvents are non-universal and a closed expression was given
for the resolvent $G_0(z,w)$ for an arbitrary potential.
Hence when switching from the canonical to the RTE or micro-canonical 
ensemble,
the delta-function constraint destroys the macroscopic universality.

A different behavior appears in the study of correlations of eigenvalues at 
the scale of the mean level spacing $1/n$. Here we recover the sine-law of the
canonical ensemble and prove its universality for the class of RTEs with 
monomial potential. This leads us to conjecture that microscopic universality
holds also for more general RTEs. The result in the microscopic limit
is not unexpected since a global constraint should not influence the local 
statistics of eigenvalues.

\indent

\underline{Acknowledgments}: 
We wish to thank G. Cicuta and L. Molinari for their
very enjoyable collaboration and many discussions on work done prior to 
this publication. Furthermore one of us (G.A.) wishes to thank the Physics 
Department of Parma for its hospitality extended to him on several occasions.
The work of G.V. is supported in part by MURST 
within the project of ``Theoretical Physics of fundamental Interactions''.

\begin{appendix}

\sect{The functions $\dx,\ \dy$ and $\dial$}\label{detxya}

In this Appendix we apply the operator $\dW(p)$ 
eq. (\ref{dWsum}) to the boundary 
conditions eqs. (\ref{bc1}) and (\ref{bc2}) and solve the linear set of
equations for the quantities $\dx,\ \dy$ and $\dial$. Using the identity
\beq
\parW (p) W^{\prime}(\om) \ =\ \frac{-1}{(p-\om)^2} \ \ ,
\label{pardV}
\eeq
we obtain 
\beqn
0 &=& \partial_p \phi(p) +\frac{1}{2}\left( M_1\dx +J_1\dy \right)
      + \cI\Vp(\om)\phi(\om)\dial \nonumber
\eeqn
\newpage
\beqn
0 &=& \partial_p (p\phi(p)) +\frac{1}{2}\left( xM_1\dx +yJ_1\dy
        \right)\nonumber\\
    &&+ \cI \om\Vp(\om)\phi(\om)\dial \ ,
\label{dbc1}
\eeqn
from eq. (\ref{bc1}) and
\beqn
0 &=& \frac{C}{2}\left(M_1\dx -\frac{\phi(p)}{p-x}\right) 
    + \frac{D}{2}\left( J_1\dx -\frac{\phi(p)}{p-y}\right)\nonumber\\
    &&+\ \dial \int_y^x\!\frac{d\lam}{2\pi}V(\lam)\sqrt{(x-\lam)(\lam-y)}
        \cI\frac{\Vp(\om)}{\om-\lam}\phi(\om) \nonumber\\
    &&+\ \frac{1}{2}\cI \frac{\Vp(\om)}{p-\om}
           \sqrt{\frac{(\om-x)(\om-y)}{(p-x)(p-y)}} \ ,
\label{dbc2}
\eeqn
from eq. (\ref{bc2}) after some calculation. Here we have introduced 
\beqn
C &\equiv& 
\int_y^x\!\frac{d\lam}{2\pi}\frac{V(\lam)}{x-\lam}\sqrt{(x-\lam)(\lam-y)} 
\ , \nonumber\\
D &\equiv&
\int_y^x\!\frac{d\lam}{2\pi}\frac{V(\lam)}{y-\lam}\sqrt{(x-\lam)(\lam-y)} 
\ .
\eeqn
We note that special care has to be taken when applying
$\dW(p)$ to $\M(\lam)$ which is then no longer an 
analytic function in $\lam$. Therefore the contour ${\cal C}_\infty$ 
has to be deformed in eq. (\ref{M}) to contain only the pole and cut in the  
integrand, and not the new pole introduced by $\dW(p)$  
(see also Appendix A in \cite{A96}).
The linear set of equations eq. (\ref{dbc1}) and
eq. (\ref{dbc2}) simplifies considerably when setting
$W\equiv 0$ because of $\Ve(\om)=\ial V(\om)$ in that case. It is this 
limit which we will need in order give a closed final expression
for the planar 2-point resolvent of the pure delta-measure without the 
auxiliary potential in eq. (\ref{Zdelta}). Using eq. (\ref{bc1}) it
follows 
\beqn
0 &=& M_1\dx -\frac{\phi(p)}{p-x}\ +\ J_1\dx -\frac{\phi(p)}{p-y}\nonumber\\
0 &=& \frac{x}{2}\left(M_1\dx -\frac{\phi(p)}{p-x}\right) 
    + \frac{y}{2}\left( J_1\dx -\frac{\phi(p)}{p-y}\right)
    + \frac{2}{\ial}\dial\nonumber\\
0 &=& \frac{C}{2}\left(M_1\dx -\frac{\phi(p)}{p-x}\right) 
    + \frac{D}{2}\left( J_1\dx -\frac{\phi(p)}{p-y}\right)\nonumber\\
    &&+ \frac{1}{\ial}A^2\dial + \frac{1}{\ial}(G_0(p)-\phi(p)) \ .
\label{dbc}
\eeqn
This can be easily solved for the desired quantities. We obtain
\beqn
M_1\dx &=& \frac{\phi(p)}{p-x} \ +\ \frac{4}{B(x-y)}(G_0(p)-\phi(p))\nonumber\\
J_1\dy &=& \frac{\phi(p)}{p-y} \ -\ \frac{4}{B(x-y)}(G_0(p)-\phi(p))\nonumber\\
\frac{1}{\ial}\dial &=& -\frac{1}{B}(G_0(p)-\phi(p)) \ ,
\eeqn
as given in eq. (\ref{dxya}) where the following 
abbreviation has been introduced
\beq
B \ =\ \ial \left(A^2\ +\  2\ \frac{C+D}{x-y}\right) \ .
\label{Bapp}
\eeq
One can easily convince oneself that it equals the form given in
eq. (\ref{B}).

\sect{RTE {\it via} inverse Laplace transform} 
\label{appb}
In this appendix we derive eq. (\ref{mic10}) which expresses expectation 
values with respect to the delta-measure in terms of averages with respect
to the canonical measure eq. (\ref{mic1}) using the inverse 
Laplace transform.
Let $F_k(M;{\bf \lam})$ be a function of the Hermitian $n \times n$
matrix $M$  and of the set of parameters 
${\bf \lam}=(\lam_1,\ldots, \lam_k)$, such that it satisfies the 
homogeneity property
\beq
\label{appb1}
F_k(tM;{\bf \lam})=t^{a}F_k(M;t^{b}{\bf \lam})
\eeq
for some real $t$, $a$ and $b$. In other words $F_k$ is a homogeneous
function  of degree $a$ with respect to matrix elements $M_{ij}$ and of
degree  $(-b)$ with respect to each parameter $\lam_i$. 
An example for such a function is the operator inside the average
of eq. (\ref{mic5}). The matrix integral
considered  in this appendix is:
\beq
\label{appb2}
I[F_k]=\int \dM \; \delta \left( A^2-\frac{1}{n} \Tr[M^{2p}] \right) \, 
 F_k(M;{\bf \lam}) \ ,
\eeq
with $p$ an integer number. It is proportional to the average 
$\langle F_k \rangle_{\delta}$.
Introducing the complex representation of
the  delta function $2 \pi \delta(x)=\int\! dy \exp [i x y]$ and then
scaling all matrix elements by a factor $(gn^2/(iy+0^{+}))^{1/2p}$, 
eq. (\ref{appb2}) reads:
\beq
I[F_k]=\int_{-\infty}^{+\infty} \frac{dy}{2 \pi} e^{i y A^2} \!\left( 
\frac{gn^2}{i y+0^{+}} 
\right)^{\frac{n^2}{2p}} \!\!\int \!\dM \; e^{-ng \Tr[M^{2p}]} F_k( \left( 
\frac{gn^2}{iy+0^{+}} \right)^{\frac{1}{2p}} \!\!\!\!M;{\bf \lam}) 
\eeq   
where we have interchanged integrals.
By using the homogeneity property (\ref{appb1}) 
we can rewrite the matrix integral as a canonical ensemble average,
up to the normalization
factor  ${\cal Z}$ eq. (\ref{mic1}):
\beqn
I[F_k] &=& {\cal Z} \int_{-\infty}^{+\infty}
 \frac{dy}{2 \pi} e^{i y A^2} \left(\frac{gn^2}{i y+0^{+}}
\right)^{\frac{n^2+a}{2p}} \langle F_k(M;\left( \frac{gn^2}{i y+0^{+}} 
\right)^{\frac{b}{2p}} {\bf \lam} ) \rangle \nonumber\\
\ &=& {\cal Z} \ (gn^2)^{\frac{n^2+a}{2p}}   \; {\cal L}^{-1} \left[  
 \frac{\langle F_k(M;\left( \frac{gn^2}{t} \right)^{\frac{b}{2p}} {\bf
     \lam}) 
\rangle}{t^{\frac{n^2+a}{2p}}} \right](A^2) \label{appbquasi} \ .
\eeqn
In the last step we used the complex representation of the
inverse  Laplace transform, i.e.
\beq 
\label{Ldef}
{\cal L}^{-1}[h(t)](x)=\frac{1}{2 \pi i} 
\int_{-i\infty+0^{+}}^{+i\infty+0^{+}} dt \; e^{t x} 
h(t) \ , \\ \ x>0 \ \ . 
\eeq
In order to obtain the correct normalization of the delta-average we evaluate
eq. (\ref{appbquasi}) in the case $F=1$ (with $a=b=0$):
\beq
\label{appbnrmlzn}
I[1]=  (gn^2)^{\frac{n^2}{2p}}  \frac{A^{\frac{n^2}{p}-2} } {\Gamma \left(
    \frac{n^2}{2p} \right) } {\cal Z} \ \ ,
\eeq
where we used the formula   
\beq
\label{useful}
{\cal L}^{-1} \left[ \frac{1}{t^{\gamma+1}} \right](x)=
\frac{x^{\gamma}}{\Gamma(\gamma+1)}\theta(x) \ \ , \ \mbox{Re}(\gamma)>-1\ \ . 
\eeq
Finally, by putting together eqs. (\ref{appbquasi}) and (\ref{appbnrmlzn}),
the ensemble average $\langle F_k \rangle_{\delta} \equiv I[F_k]/I[1]
$ with  respect to the measure $\delta(A^2-\Tr [M^{2p}]/n)$ reads:
\beq
\label{formula}
\langle F_k \rangle_{\delta}= \frac{(gn^2)^{\frac{a}{2p}} }{ 
A^{\frac{n^2}{p}-2}} 
\Gamma\left( \frac{n^2}{2p} \right) {\cal L}^{-1} \left[  
 \frac{\langle F_k(M;\left( \frac{gn^2}{t} \right)^{\frac{b}{2p}} {\bf
     \lam}) 
\rangle}{t^{\frac{n^2+a}{2p}}} \right] (A^2) \ \ ,
\eeq
which is just eq. (\ref{mic10}).

\end{appendix}

\newpage

\newcommand{\NP}[3]{{\it Nucl. Phys. }{\bf B#1} (#2) #3}
\newcommand{\PL}[3]{{\it Phys. Lett. }{\bf B#1} (#2) #3}
\newcommand{\PR}[3]{{\it Phys. Rev. }{\bf #1} (#2) #3}
\newcommand{\PRE}[3]{{\it Phys. Rev. }{\bf E#1} (#2) #3}
\newcommand{\IMP}[3]{{\it Int. J. Mod. Phys }{\bf #1} (#2) #3}
\newcommand{\MPL}[3]{{\it Mod. Phys. Lett. }{\bf #1} (#2) #3}


\begin{thebibliography}{99}

\bibitem {guhr} T. Guhr, A. M\"{u}ller-Groeling and H.A.
Weidenm\"{u}ller, {\it Phys. Rep.} {\bf 299} (1998) 190.

\bibitem{Been} C.W. Beenakker, {\it Rev. Mod. Phys.} {\bf 69} (1997) 731.

\bibitem{AJM}  J. Ambj{\o}rn, J. Jurkiewicz and Y. Makeenko,
           \PL{251}{1990}{517}.

\bibitem{AA} G. Akemann and J. Ambj{\o}rn,
{\it J. Phys.} {\bf A29} (1996) L555.

\bibitem{BZ} E. Br\'ezin and A. Zee, \NP{402}{1993}{613}.

\bibitem{ADMN} G. Akemann, P. H. Damgaard, U. Magnea and S. Nishigaki,
\NP{487}{1997}{721}.

\bibitem{BB} M.J. Bowick and E. Br\'ezin, \PL{268}{1991}{21}.
\bibitem{KF} E. Kanzieper and V. Freilikher, \PR{E55}{1997}{3712}.

\bibitem{ACMVII} G. Akemann, G.M. Cicuta, L. Molinari and G. Vernizzi,
\PRE{60}{1999}{5287}.

\bibitem{Pastur} A. Khorunzhy, B. Khoruzhenko and L. Pastur, {\it J. Phys.} 
{\bf A28} (1995) L31; {\it J. Math. Phys.} {\bf 37} (1996) 5033.

\bibitem{danna} J. D'Anna and A. Zee, \PRE{53}{1996}{1399}.

\bibitem{Iso}S. Iso and A. Kavalov, \NP{501}{1997}{670}.

\bibitem{das} S.R. Das, A. Dhar, A.M. Sengupta and S.R. Wadia, 
\MPL{A5}{1990}{1041}. 

\bibitem{CM}G.M. Cicuta and E. Montaldi, \MPL{A5}{1990}{1927}.

\bibitem{met}  
N. Rosenzweig, {\it Statistical Physics}, Brandeis Summer
Institute 1962, G. Uhlenbeck et al., W.A. Benjamin Inc. 1963.

\bibitem{Bronk}
B.V. Bronk, Princeton University thesis 1964, as cited 
in \cite{Mehta}, Chapter 19.

\bibitem{korchemsky} G.P. Korchemsky, \MPL{A7}{1992}{3081}.

\bibitem{ACMV} G. Akemann, G.M. Cicuta, L. Molinari and G. Vernizzi,
\PRE{59}{1999}{1489}.

\bibitem{Ambbook} J. Ambj{\o}rn, B. Durhuus and T. Jonsson, 
{\it Quantum Geometry}, Cambridge University Press 1997.

\bibitem{Gra} G. Vernizzi, {\it Classes of Universality in Random Matrix 
Theory}, PhD-thesis, Parma 1999.

\bibitem{SV}E.V. Shuryak and J.J.M. Verbaarschot, {\it Nucl. Phys.} {\bf A560}
(1993) 306. 

\bibitem{AMB93} J. Ambj{\o}rn, L. Chekhov, C. F. Kristjansen and Yu. Makeenko,
\NP{404}{1993}{127}.

\bibitem{A96} G. Akemann \NP{482}{1996}{403}.

\bibitem{DJ} F. David, \NP{348}{1991}{507}. 

\bibitem{Ju} J. Jurkiewicz, \PL{245}{1990}{178}.

\bibitem{Itoi}C. Itoi, \NP{493}{1997}{651}.

\bibitem{AKM92} J. Ambj{\o}rn, C.F. Kristjansen and Yu. Makeenko,
\MPL{A7}{1992}{3187}. 

\bibitem{A97}G. Akemann, \NP{507}{1997}{475}.

\bibitem{Delannay} R. Delannay and G. Le Ca\"{e}r, {\it J. Phys.} 
{\bf A33} (2000) 2611. 

\bibitem{Mehta} M.L. Mehta, {\it Random Matrices}, Academic Press 1991.


\end{thebibliography}
\end{document}